\documentclass[journal=jacsat,manuscript=article]{achemso}
\usepackage{mathrsfs}
\usepackage{amsmath}% needed for subequations
\usepackage{amsfonts}
\usepackage{mathrsfs}
\usepackage{amsmath}% needed for subequations
\usepackage{color}
\usepackage{natbib}
\usepackage{textcomp}
\usepackage{graphicx}
\usepackage{bm}% bold maths
\usepackage{amssymb}
\usepackage{xspace}
\usepackage{epstopdf}
\usepackage{dcolumn}% Align table columns on decimal point
\usepackage{longtable}
\usepackage{multirow}
\usepackage[colorlinks=true, letterpaper=true, pdfstartview=FitV, linkcolor=blue, citecolor=blue, urlcolor=blue]{hyperref}
\usepackage{float}
\usepackage{chemformula} % Formula subscripts using \ch{}
\usepackage[T1]{fontenc} % Use modern font encodings

\author{Jing-Yang You}
\affiliation{School of Physical Sciences,University of Chinese Academy of Sciences, Beijing 100049, China}
\author{Bo Gu}
\email{gubo@ucas.ac.cn}
\affiliation{Kavli Institute for Theoretical Sciences, and CAS Center for Excellence in Topological Quantum Computation, University of Chinese Academy of Sciences, Beijing 100190, China}
\alsoaffiliation{Physical Science Laboratory, Huairou National Comprehensive Science Center, Beijing 101400, China}
\author{Gang Su}
\email{gsu@ucas.ac.cn}
\alsoaffiliation{School of Physical Sciences,University of Chinese Academy of Sciences, Beijing 100049, China}
\alsoaffiliation{Kavli Institute for Theoretical Sciences, and CAS Center for Excellence in Topological Quantum Computation, University of Chinese Academy of Sciences, Beijing 100190, China}
\alsoaffiliation{Physical Science Laboratory, Huairou National Comprehensive Science Center, Beijing 101400, China}

\title{Flat Band and Hole-induced Ferromagnetism in a Novel Carbon Monolayer}

\begin{document}

\begin{abstract}
In recent experiments, superconductivity and correlated insulating states were observed in twisted bilayer graphene (TBG) with small magic angles, which highlights the importance of the flat bands near Fermi energy. However, the moir\'{e} pattern of TBG consists of more than \textit{ten thousand} carbon atoms that is not easy to handle with conventional methods. By density functional theory calculations, we obtain a flat band at E$_F$ in a novel carbon monolayer coined as cyclicgraphdiyne with the unit cell of \textit{eighteen} atoms. By doping holes into cyclicgraphdiyne to make the flat band partially occupied, we find that cyclicgraphdiyne with 1/8, 1/4, 3/8 and 1/2 hole doping concentration shows ferromagnetism (half-metal) while the case without doping is nonmagnetic, indicating a hole-induced nonmagnetic-ferromagnetic transition. The calculated conductivity of cyclicgraphdiyne with 1/8, 1/4 and 3/8 hole doping concentration is much higher than that without doping or with 1/2 hole doping. These results make cyclicgraphdiyne really attractive. By studying several carbon monolayers, we find that a perfect flat band may occur in the lattices with both separated or corner-connected triangular motifs with only including nearest-neighboring hopping of electrons, and the dispersion of flat band can be tuned by next-nearest-neighboring hopping. Our results shed insightful light on the formation of flat band in TBG. The present study also poses an alternative way to manipulate magnetism through doping flat band in carbon materials.
\end{abstract}

\maketitle

%%%%%%% Main text %%%%%%%%%%%%%%%%%%%%%

\section{Introduction}
A twisted bilayer graphene (TBG) with small angles was predicted to have moir\'{e} pattern with a very large unit cell of more than ten thousand carbon atoms. At the magic angles, the two layers become strongly coupled, and a flat band appears~\cite{Bistritzer2011}. A tight-binding model~\cite{Morell2010} and a continuum model of TBG~\cite{Santos2012} were thus studied, and the confined states were also proposed in TBG~\cite{Laissardiere2012}. Recently, the superconductivity and correlated insulating states at half-filling were observed in TBG, which highlight the importance of the flat bands near Fermi energy~\cite{Cao2018,Cao2018a}. The strong coupling phases in TBG (e.g.~\cite{Kang2019}) were explored. It was shown that the degeneracy of the low-energy flat bands can be lifted by the electric fields~\cite{Chebrolu2019}. The topological properties of electronic band structure~\cite{Koshino2019,Po2019}, ferromagnetic Mott state~\cite{Seo2019} and metal-insulator transition~\cite{Yuan2018,Padhi2019},  extended Hubbard model, nematic superconductivity by density wave fluctuations~\cite{Koshino2018,Isobe2018,Kozii2019} in TBG were addressed.

The study of TBG has inspired extensive studies in other twisted bilayer materials. Moir\'{e} excitons, moir\'{e} trapped valley excitons, and resonantly hybridized excitons in van der Waals heterostructures were observed in experiments~\cite{Tran2019,Seyler2019,Jin2019,Alexeev2019}. Van der Waals crystals with discretized Eshelby twist in two-dimensional (2D) materials~\cite{Liu2019}, and van der Waals contacts between three-dimensional metals and 2D semiconductors~\cite{Wang2019} were investigated. Chiral twisted van der Waals nanowires~\cite{Sutter2019}, and helical van der Waals crystals~\cite{Liu2019a} were also studied.

In this paper, by means of the first-principles calculations, we find that the flat band near the E$_F$ could appear in a carbon monolayer with unit cell of eighteen atoms, which is coined cyclicgraphdiyne. By doping holes into cyclicgraphdiyne to make the flat band partially occupied, cyclicgraphdiyne with 1/8, 1/4, 1/8 and 1/2 hole doping concentration becomes ferromagnetic, while the case without doping holes remains nonmagnetic, showing a hole doping induced nonmagnetic-ferromagnetic transition in carbon materials. After carefully studying the structural stability, we disclose that cyclicgraphdiyne is viable. By investigating several carbon monolayers, we notice that a perfect flat band can occur in the lattices with both separated or corner-connected triangular motifs and nearest-neighboring hopping of electrons, and the dispersion of flat band can be tuned by next nearest-neighboring hopping. Our results shed light on the formation reasoning of flat band in TBG. The present study also provides a simple 2D platform of carbon atoms to probe the flat bands and magnetism in 2D systems.

\section{Results}
\subsection{Structure and Stability}

\begin{figure}[tbhp]
  \centering
  \includegraphics[scale=0.9,angle=0]{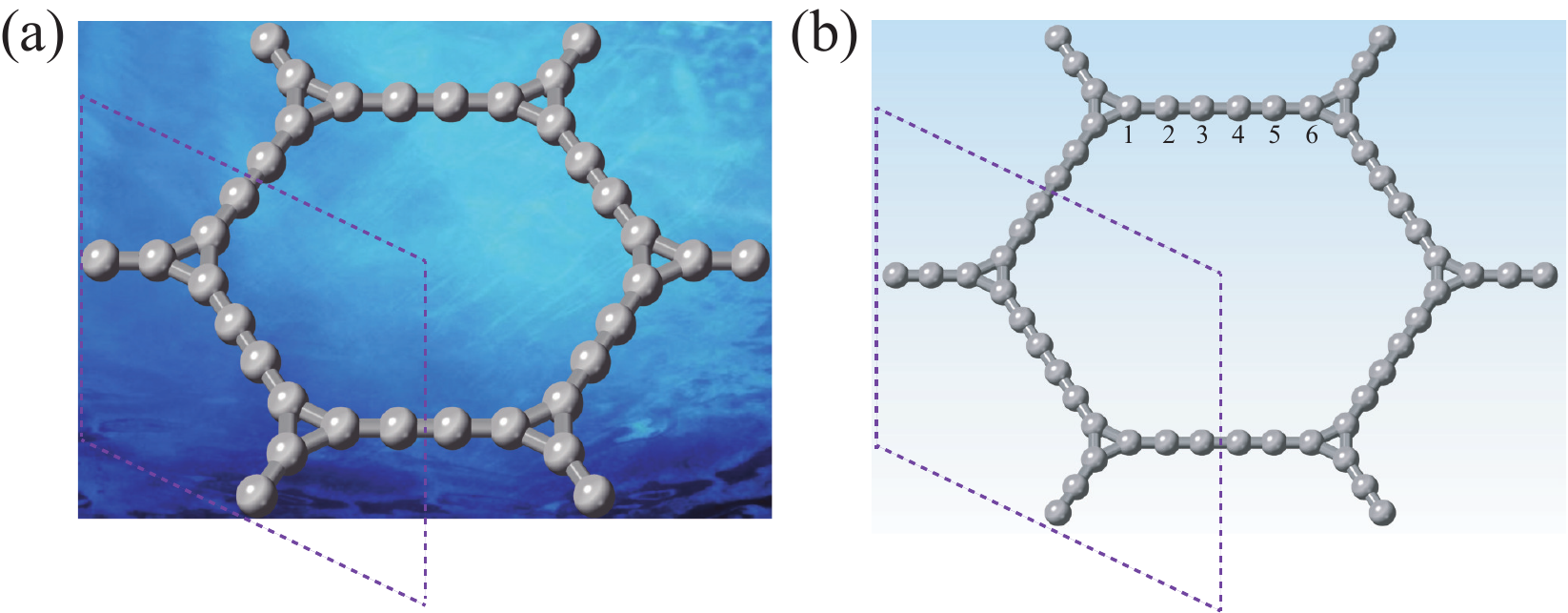}\\
  \caption{The crystal structures of (a) cyclicgraphyne with unit cell of twelve carbon atoms and (b) cyclicgraphdiyne with unit cell of eighteen carbon atoms as indicated by dashed lines, as well as the atom index labeled in the chain between two triangles in cyclicgraphdiyne. Images of (a) and (b) created using VESTA 3.4.7 software (jp-minerals.org).}\label{fig1}
\end{figure}

Cyclicgraphyne and cyclicgraphdiyne can be made by inserting an acetylene linkage and two acetylene linkages between two neighboring carbon triangles~\cite{Zhao2013}, respectively, as shown in Fig.~\ref{fig1}. The crystal structure has the space group $P6/mmm$ (No.191). Our calculations show that the equilibrium lattice constants of cyclicgraphyne and cyclicgraphdiyne are $a_0$ = 9.6452 {\AA} and $a_0$ =14.0920 {\AA}, respectively. To unveil the stabilities of cyclicgraphyne and cyclicgraphdiyne, the phonon spectra were calculated, as shown in Figs. S1(a) and (b) (supplemental material). It can be seen that there is no negative frequency phonon in the whole Brillouin zone, indicating that cyclicgraphyne and cyclicgraphdiyne are kinetically stable. To further examine the thermal stability, we performed molecular dynamics simulations by considering $3\times3\times1$ supercells of cyclicgraphyne and cyclicgraphdiyne with 108 and 162 carbon atoms, respectively. As displayed in Fig. S1(c), after being heated at 500 K, 700 K, and 1000 K for 6 ps with a time step of 3 fs, no structural changes occur, indicating that cyclicgraphyne and cyclicgraphdiyne are also dynamically stable. As shown in Fig. S1(d), we present the free energy of cyclicgraphyne and cyclicgraphdiyne, compared with three experimentally available carbon allotropes: T-carbon~\cite{Sheng2011}, graphene~\cite{Novoselov2004}, and graphdiyne~\cite{Li2010}. The results suggest that they may be feasible for experimental synthesis.

\subsection{Flat Band in Cyclicgraphyne and Cyclicgraphdiyne}

\begin{figure}[!htp]
  \centering
  \includegraphics[scale=0.81,angle=0]{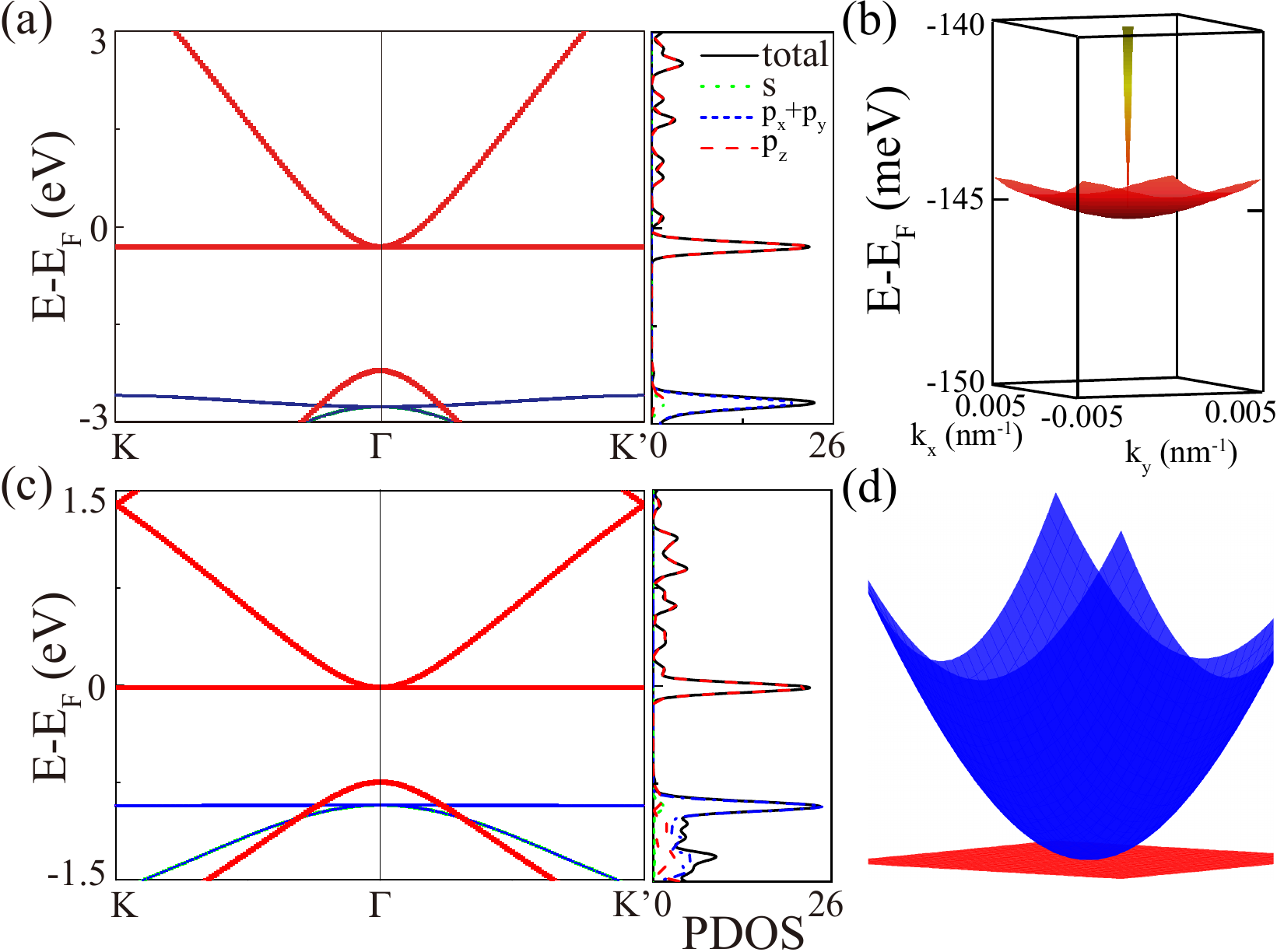}\\
  \caption{(a) The electronic band structure of cyclicgraphyne, where below 145 meV from $E_F$, a flat band (b) in the whole momentum space with bandwidth of about 2 meV is obtained by DFT calculations. (c) The electronic band structure of cyclicgraphdiyne, where a perfect flat band near $E_F$ is clearly observed according to DFT calculations. (d) The three-dimensional plot of (c) near Fermi level in momentum space.}\label{fig2}
\end{figure}

By DFT calculations, the band structure of cyclicgraphyne near Fermi energy $E_F$ are shown in Fig.~\ref{fig2}(a). At 145 meV below Fermi energy $E_F$, a flat band with bandwidth of about 2 meV is observed [Fig.~\ref{fig2}(b)]. The flat band is mainly coming from $p_z$ electrons, as demonstrated in the partial density of state (PDOS) in Fig.~\ref{fig2}(a). Our result of flat band in cyclicgraphyne is consistent with previous studies~\cite{Chen2018,Mi2019}. As shown in Fig.~\ref{fig2}(c), a very similar electronic band structure is obtained in cyclicgraphdiyne, where a perfect flat band appears nearly at $E_F$ (see also Fig.~\ref{fig2}(d)).

\subsection{Origin of Flat Band in Cyclicgraphdiyne}

\begin{figure}[tbhp]
  \centering
  \includegraphics[scale=1.1,angle=0]{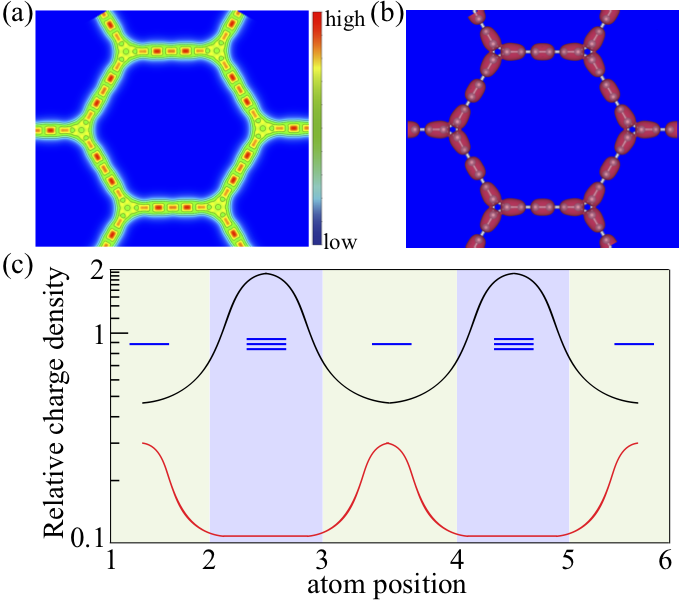}\\
  \caption{(a) Total electron charge density of cyclicgraphdiyne. (b) Electron charge density of the flat band near Fermi level. Images of (a) and (b) created using VESTA 3.4.7 software (jp-minerals.org). (c) A schematic logarithmic plot of the relative value of the total electron charge density (black line) to the electron charge density of the flat band (red line) along the chain between two neighboring carbon triangles. The blue lines represent the triple and single bonds. Note that the peaks of the total electron charge density appear at the triple bonds, while the peaks of the electron charge density of the flat band appear at the single bonds and triangles. The peak of the electron charge density of the flat band is about 2 percent of the peak of the total charge density.}\label{fig3}
\end{figure}

The total electron charge density of cyclicgraphdiyne is illustrated in Fig.~\ref{fig3}(a). It is instructive to note that the peak of total electron charge density occurs at acetylene linkages, while the valley appears at single bonds and triangles. If we fix the energy at the flat band near E$_F$, the obtained electron charge density of the flat band is shown in Fig.~\ref{fig3}(b). It has a characteristic of $p_z$ orbitals, and is not distributed on the carbon atoms at acetylene linkages in cyclicgraphdiyne. By comparing Figs.~\ref{fig3}(a) and (b), we know that the flat band should mainly come from the valley part of the total electron charge density as shown in Fig.~\ref{fig3}(c).

\subsection{Hole-induced ferromagnetism in Cyclicgraphdiyne}

\begin{figure}[tbhp]
  \centering
  \includegraphics[scale=0.7,angle=0]{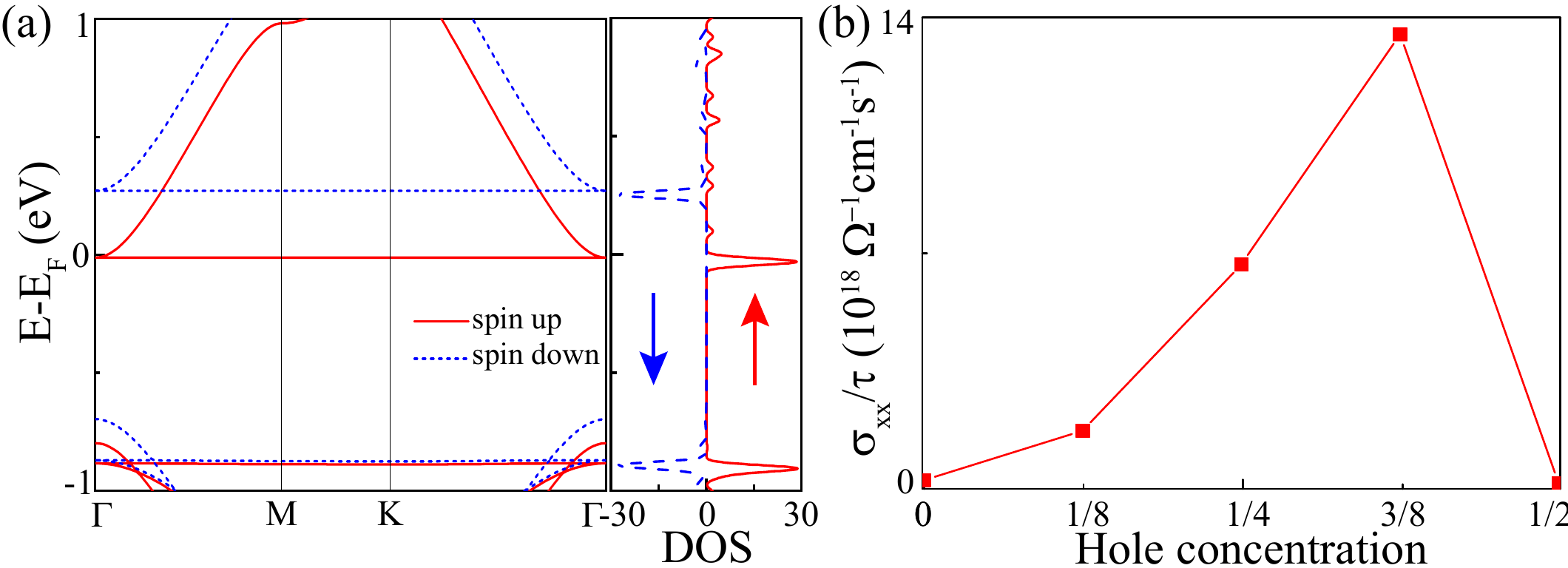}\\
  \caption{(a) The band structure and density of states (DOS) for spin up and spin down electrons of cyclicgraphdiyne with one hole doping in the unit cell (i.e. 1/2 hole concentration), showing a ferromagnetic ground state (a half-metallic state). (b) The conductivity ($\sigma$) over the relaxation time ($\tau$) against the hole doping concentration in 2$\times$2$\times$1 supercell of cyclicgraphdiyne.}\label{fig4}
\end{figure}

The interplay between flat band and Coulomb interaction may cause the flat band to be magnetic. To investigate the flat band magnetism, we doped a hole in the unit cell of cyclicgraphdiyne to make the flat band near E$_F$ half filled. Figure~\ref{fig4}(a) shows the band structure of the doped ferromagnetic ground state with an average spin moment about 0.03 $\mu_B$ per atom. The spin splitting of the flat band is about 0.28 eV. To study the change of conductivity with different hole doping in cyclicgraphdiyne, we calculated the conductivity divided by the relaxation time [Fig.~\ref{fig4}(b)] in a 2$\times$2$\times$1 supercell of cyclicgraphdiyne with 0, 1, 2, 3 and 4 hole dopants, respectively, i.e. the hole doping concentration corresponds to the hole filling of 0, 1/8, 1/4, 3/8 and 1/2, respectively. From Fig.~\ref{fig4}(b), we note that the conductivity in half and full fillings of the flat band in cyclicgraphdiyne is much smaller than that in the case of 1/8, 1/4 and 3/8 hole doping. The behavior of conductivity in Fig.~\ref{fig4}(b) can be understood in terms of the density of states (DOS). The DOS near Fermi energy with hole filling of 1/8, 1/4, 3/8 and 1/2 is presented in Fig. S2. It is shown that the DOS at Fermi energy becomes finite and appears only for single species electrons with hole hoping, and becomes to nearly zero at half filling. This result reveals that the hole-doped cyclicgraphdiyne is a half-metal, which would be useful for spintronics. The resemblance of conductivity as a function of hole doping between cyclicraphdiyne in Fig.~\ref{fig4}(b) and TBG observed in the experiment~\cite{Cao2018,Cao2018a} makes cyclicgraphdiyne really attractive.

\subsection{Flat Band and Tight-Binding Model}

\begin{figure}[!hbt]
  \centering
  \includegraphics[scale=0.6,angle=0]{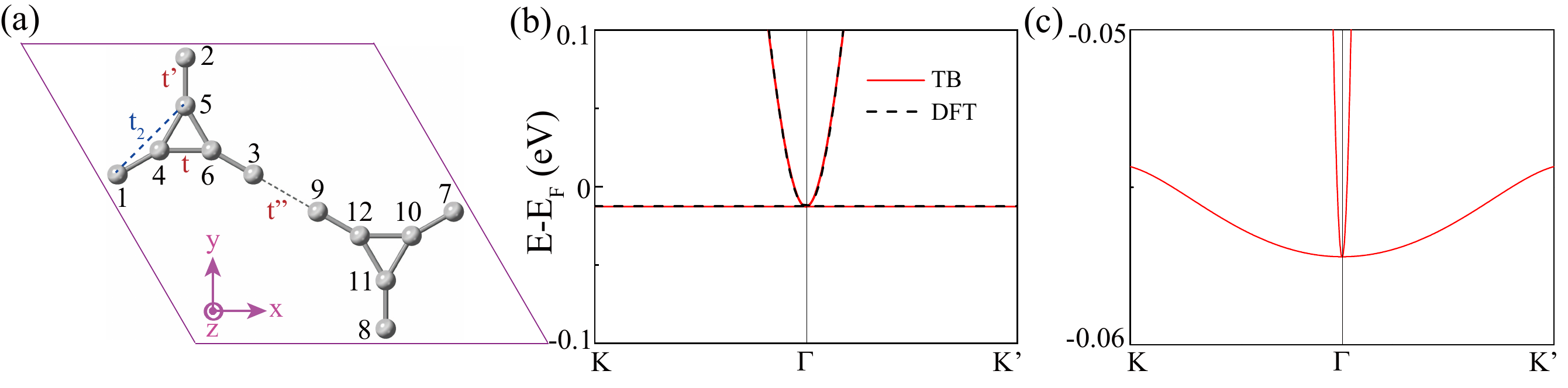}\\
  \caption{(a) The crystal structures of cyclicgraphdiyne with atom index. (b) The tight-binding electronic bands based on $p_z$ orbitals of carbon atoms for cyclicgraphdiyne with hop
  ping parameters $t$ = -2.76, $t^{'}$ = -1.99 and $t^{''}$ = -1.42. (c) The tight-binding electronic bands with a small NNN hopping $t_2=0.05$
  based on (b).}\label{fig5}
\end{figure}

To understand the origin of the flat band in cyclicgraphdiyne, we construct a specified tight-binding model based on the $p_z$ orbitals of cyclicgraphdiyne. A simplified effective model based on $p_z$ orbitals of carbon atoms for cyclicgraphdiyne can be written by a 12-bands Hamiltonian, where we only consider the carbon atoms on the triangles and their nearest neighboring carbon atoms, and the interaction between the carbon atoms on the acetylene linkages is summarized to the effective hopping $t^{''}$ as shown in Fig.~\ref{fig5}. So we have
\begin{eqnarray*}\label{Det}
H(k_x,k_y)=
\begin{pmatrix}
0 & B &C &0\\
B^{\ast} & A &0 &0\\
C^{\ast} &0  &0 &B^{\ast}\\
0 &0 &B &A^{\ast}
\end{pmatrix}
,
\end{eqnarray*}
where
\begin{eqnarray*}
A=
\begin{pmatrix}
0 &a &b \\
a^{\ast} &0 &c\\
b^{\ast} &c{\ast} &0
\end{pmatrix}
,
\end{eqnarray*}
$a=te^{i\cdot(k_x/2+\sqrt{3}k_y/2)\cdot l} $, $b=te^{i\cdot k_x\cdot l}$, $c=te^{i\cdot(k_x/2-\sqrt{3}k_y/2)\cdot l}$,
$B=Diag[$  $t^{'}e^{i\cdot0.5473\cdot(3k_x/2+\sqrt{3}k_y/2)\cdot l}, t^{'}e^{-i\cdot0.5473\cdot\sqrt{3}k_y\cdot l},$ $t^{'}e^{-i\cdot0.5473\cdot(3k_x/2-\sqrt{3}k_y/2)\cdot l}]$, $C=Diag[t^{''}e^{i\cdot1.5476\cdot(3k_x/2+\sqrt{3}k_y/2)\cdot l}, t^{''}e^{-i\cdot1.5476\cdot\sqrt{3}k_y\cdot l}, t^{''}e^{-i\cdot1.5476\cdot(3k_x/2-\sqrt{3}k_y/2)\cdot l}]$, $l$=1.425 \AA    (the bond length of two carbon atoms at triangles),
$t$, $t^{'}$ and $t^{''}$ are the nearest-neighbouring (NN) hopping parameters as shown in Fig.~\ref{fig5}(a). By DFT calculations~\cite{Marzari1997}, these NN hopping parameters were obtained, and the tight-binding (TB) flat band fits well with the DFT bands as shown in Fig.~\ref{fig5}(b). If the next-nearest-neighbouring (NNN) hopping (such as $t_2$) is included, which is described by
\begin{eqnarray*}\label{Det}
H_{NNN}(k_x,k_y)=t_2
\begin{pmatrix}
0 & D &0 &0\\
D^{\ast} & 0 &0 &0\\
0 &0  &0 &D^{\ast}\\
0 &0 &D &0
\end{pmatrix}
,
\end{eqnarray*}
where
\begin{eqnarray*}
D=
\begin{pmatrix}
0 &e &f \\
m &0 &n\\
p &q &0
\end{pmatrix}
,
\end{eqnarray*}
with $e=e^{i[(0.5473\cdot3+1)k_x/2+(0.5473+1)\sqrt{3}k_y/2]\cdot l}$, $f=e^{i[(0.5473\cdot3+2)k_x/2+0.5473\sqrt{3}k_y/2)]\cdot l}$, $m=e^{i[-k_x/2-(0.5473\cdot2+1)\sqrt{3}k_y/2]\cdot l}$, $n=e^{i[k_x/2-(0.5473\cdot2+1)\sqrt{3}k_y/2]\cdot l}$, $p=e^{i[-(0.5473\cdot3+2)k_x/2+0.5473\sqrt{3}k_y/2)]\cdot l}$, $q=e^{i[-(0.5473\cdot3+1)k_x/2+(0.5473+1)\sqrt{3}k_y/2]\cdot l}$, the obtained perfect flat band can be tuned. A small NNN hopping $t_2=-0.05$ can induce a tiny bandwidth as shown in Fig.~\ref{fig5}(c).

The above results are usually implemented in a kagome lattice, where a perfect flat band appears with only NN hopping and can be tuned out by NNN hopping. By comparing the results of cyclicgraphdiyne in Fig.~\ref{fig5} and those of kagome lattice in Fig. S3, and studying several materials with triangular and hexagonal structures, we can conclude that a perfect flat band can occur in the lattices with both separated or corner-connected triangular structures with only NN hopping, and can be tuned out by the NNN hopping.

\section{Discussions}

\subsection{Flat Band and Hidden Valley Kagome Lattice in TBG}

For TBG with small magic angles, the unit cell can be constructed from moir\'{e} pattern, which may be regarded as a huge triangular lattice~\cite{Cao2018}, where the lattice points are comprised of the peaks of local density of states (LDOS) of TBG, while the valleys of LDOS, i.e., the edge center of triangles, form a hidden kagome lattice as shown in Fig. S5. It is known that the kagome lattice with only NN hopping can induce the flat band~\cite{Mielke1992,Zhou2014}, as we also addressed in the Supplemental Materials. In TBG, the hidden kagome lattice obtained from moir\'{e} pattern is quite large, so the NNN hopping should be much smaller than the NN hopping, and a nearly flat band can be obtained from such a hidden kagome moire lattice. Analog to the flat band in cyclicgraphdiyne [Fig.~\ref{fig2}(d)], where the flat band mainly comes from the valleys of the total electron charge density, we believe that the flat band in TBG may be attributed to the hidden kagome lattice in moir\'{e} pattern.

\section{Conclusions}
In summary, by density functional theory calculations, we propose a novel carbon monolayer with unit cell of eighteen atoms, coined as cyclicgraphdiyne. A flat band appears at E$_F$ in cyclicgraphdiyne. By doping holes to make the flat band in cyclicgraphdiyne partially occupied, we find that cyclicgraphdiyne with 1/8, 1/4, 3/8 and 1/2 hole doping concentration becomes ferromagnetic, and the conductivity of cyclicgraphdiyne with 1/8, 1/4 and 3/8 hole doping concentration is much higher than the case that the flat band in cyclicgraphdiyne is full or half filled. These results make the cyclicgraphdiyne proposed here very attractive. By exploring different carbon monolayers, we show that a perfect flat band could appear in the lattices with both separated or corner-connected triangular structures and only NN hopping, while the flat band can be tuned out by NNN hopping. Owing to the resemblance of the electronic structures between cyclicgraphdiyne and TBG, our results may also shed useful light on the formation reasoning of the flat band in TBG. We believe that the flat band in TBG may be attributed to a hidden valley kagome lattice in moir\'{e} pattern. This present study poses a simple carbon monolayer and gives an alternative reasoning to the formation of the flat bands and manipulating magnetism in 2D carbon materials.

\section{Computational Method}
Our first-principles calculations were based on the density-functional theory (DFT) as implemented in the Vienna \textit{ab initio} simulation package (VASP)~\cite{Kresse1996}, using the projector augmented wave method~\cite{Bloechl1994}. The generalized gradient approximation with Perdew-Burke-Ernzerhof~\cite{Perdew1996} realization was adopted for the exchange-correlation functional. The plane-wave cutoff energy was set to 550 eV. The Monkhorst-Pack $k$-point mesh~\cite{Monkhorst1976} of size $11\times11\times 1$ was used for the BZ sampling. The structure relaxation considering both the atomic positions and lattice vectors was performed by the conjugate gradient (CG) scheme until the maximum force on each atom was less than 0.0001 eV/{\AA}, and the total energy was converged to $10^{-8}$ eV with Gaussian smearing method. To avoid unnecessary interactions between the monolayer and its periodic images, the vacuum layer is set to 15 {\AA}. The crystal structures and charge density were plotted with the software VESTA~\cite{Momma2008}, and the images in the figure legend were created with the software SciDAVis~\cite{SciDAVis111}. The unit of hole-concentration is dimensionless. The 100$\%$ doping means 2 holes doped in area of 172 {\AA} (the area of unit cell of cyclicgraphdiyne). We used Hubbard $U$ (with values 1 and 4 eV) correction in DFT calculations and the electronic band structure was found unaltered. The conductivity is calculated with the package BoltzTrap~\cite{Madsen2006} based on a smoothed Fourier interpolation of the bands at temperature of 10 K.

\section{Acknowledgements}
The authors acknowledge Zheng Zhu for valuable discussions. B.G. is supported by the National Natural Science Foundation of China (Grant No. Y81Z01A1A9), the Chinese Academy of Sciences (Grant No. Y929013EA2), the University of Chinese Academy of Sciences (Grant No. 110200M208), and the Beijing Natural Science Foundation (Grant No. Z190011). J.Y.Y. and G.S. are supported in part by the National Key R\&D Program of China (Grant No. 2018YFA0305800), the Strategic Priority Research Program of the Chinese Academy of Sciences (Grant No. XDB28000000), the National Natural Science Foundation of China (Grant No.11834014), and Beijing Municipal Science and Technology Commission (Grant No. Z118100004218001).

\end{document}